\documentclass[mathleft
]{an}
\usepackage{graphicx}

\usepackage{times}
\newcommand{\mdot}{\mbox{$\dot{M}$}}
\newcommand{\Rstar}{\mbox{$R_\ast$}}

\newcommand{\vinf}{\mbox{$v_\infty$}}
\newcommand{\xmm}{{\em XMM-Newton}}
\newcommand{\cxo}{{\em Chandra}}

\newcommand{\zoph}{\mbox{$\zeta$\,Oph}}

\newcommand{\Lbol}{\mbox{$L_{\rm bol}$}}

\newcommand{\Lx}{\mbox{$L_{\rm X}$}}

\newcommand{\Msun}{\mbox{$M_\odot$}}
\newcommand{\myr}{\mbox{$M_\odot\,{\rm yr}^{-1}$}}
\newcommand{\lsim}{\raisebox{-.4ex}{$\stackrel{<}{\scriptstyle \sim}$}}
\newcommand{\msim}{\raisebox{-.4ex}{$\stackrel{>}{\scriptstyle \sim}$}}
\newcommand{\mim}{\mbox{$\mu$m}}

\def \etal   {\hbox{et~al.\/}}

\begin{document}

\Pagespan{147}{}
\Yearpublication{2011}%
\Yearsubmission{2011}%
\Month{1}%
\Volume{332}%
\Issue{2}%
 \DOI{10.1002/asna.201111516}%

\title{First detection of a magnetic field in the fast rotating runaway Oe star
$\zeta$\,Ophiuchi\thanks{Based on observations obtained at the European Southern Observatory, 
Paranal, Chile (ESO programme 081.C-0410(A)).}}

\received{2011 Jan 21} \accepted{2011 Jan 26} \publonline{2011 Feb 15}

\author{{S.~Hubrig\inst{1}\fnmsep\thanks{Corresponding author: {shubrig@aip.de}}}
\and
L.M.~Oskinova\inst{2}
\and
M.~Sch\"oller\inst{3}
}

\titlerunning{Magnetic field in the runaway Oe star \zoph}
\authorrunning{S. Hubrig, L.M. Oskinova \& M. Sch\"oller }

\institute{
Astrophysikalisches Institut Potsdam, An der Sternwarte~16, 14482~Potsdam, Germany
\and
Universit\"at Potsdam, Institut f\"ur Physik und Astronomie, 14476~Potsdam, Germany
\and
European Southern Observatory, Karl-Schwarzschild-Str.~2, 85748~Garching, Germany
}



\abstract{%
The star $\zeta$\,Ophiuchi is one of the brightest massive stars in the northern hemisphere and 
was intensively studied in various wavelength domains. The currently available observational material
suggests that certain observed phenomena are related to the presence of a magnetic field. 
We acquired spectropolarimetric observations of $\zeta$\,Oph
with FORS\,1 mounted on the 8-m Kueyen telescope of the VLT to investigate if a magnetic field
is indeed present in this star.
Using all available absorption lines, we detect a mean longitudinal magnetic field 
$\left< B_z\right>_{\rm all}= 141\pm45$\,G, confirming the magnetic nature of this star.
We review the X-ray properties of $\zeta$\,Oph with the aim to
understand whether the X-ray emission of \zoph\ is dominated by magnetic
or by wind instability processes. 
}

\maketitle

\section{Introduction}
\label{sect:intro}

During the last years a gradually increasing number of O, early B-type, and WR stars have been investigated 
for magnetic fields, and as a result, about a dozen magnetic O-type stars  
are presently known (e.g., Hubrig et al.\ \cite{Hubrig2008}; Martins et al.\ \cite{Martins2010};
Hubrig et al.\ \cite{Hubrig2011a}).
The recent detections of magnetic fields in 
massive stars generate a strong motivation to study the correlations between 
evolutionary state, rotation velocity, and surface composition, and to
understand the origin and the role of magnetic fields in massive stars.

The star $\zeta$\,Ophiuchi (=HD\,149757) of spectral type O9.5V is a
well-known rapidly rotating runaway  star with extremely interesting
characteristics. It undergoes episodic mass loss seen as emission in
H$\alpha$, and it  is possible that it rotates with almost break-up
velocity with $v$\,sin\,$i=400$\,km\,s$^{-1}$ (Kambe  et al.\
\cite{Kambe1993}). Various studies indicate different types of spectral
and photometric variability. The UV resonance lines show multiple
discrete absorption components (DAC) in the UV (e.g.\ Howarth et al.\
\cite{Howarth1984}) and strong line profile variations in optical
spectra reconciled with traveling sectorial modes of high degree (e.g.\
Reid et al.\ \cite{Reid1993}).  Highly precise {\em MOST}
(Microvariability and Oscillations of Stars) satellite photometry in
2004 has yielded at least a dozen significant oscillation frequencies
between 1 and 10 cycles/day, hinting at a behaviour similar to 
$\beta$~Cephei-type stars (Walker et al.\ \cite{Walker2005}). No
unambiguous rotation period could be identified in spectroscopic and 
photometric observations, although Balona \& Kambe
(\cite{BalonaKambe1999})
favored a period in the region of 1 cycle/day.

$\zeta$\,Oph is also well-known for its variability in the X-ray band. 
Oskinova \etal\ (\cite{Oskinova2001}) studied the {\em ASCA} observations of
\zoph\ that covered just more than the expected rotation period of the
star. A clearly detected periodic X-ray flux variability with $\sim$20\%\
amplitude  in the {\em ASCA} passband (0.5-10\,keV) was reported. A
period of 0.77\,d was detected and a possible connection with the
recurrence time (0.875\,d$\pm$0.167\,d) of the DACs
in UV spectra of the star was discussed. The DACs in
the spectra of O stars are commonly explained by 
large-scale structures in the stellar wind, modulated by
rotation and possibly related to a surface magnetic field 
(Cranmer \& Owocki \cite{CranmerOwocki1996}).
Waldron \etal\ (in preparation, private communication) found that
{\em SUZAKU} data on \zoph\ suggest a period of
$\sim$0.98\,d that is consistent but slightly larger than
the X-ray periodicities found in {\em ASCA} data (Oskinova \etal\ \cite{Oskinova2001})
and in {\em Chandra} HETGS data (Waldron \cite{Waldron2005}).  In addition,
the HETGS data appear to indicate an additional cyclic period of
$\sim$0.33\,d in the hard X-ray band ($>$1.2\,keV).  

The results of our previous studies seem to indicate that the presence
of a magnetic field  is more frequently detected in candidate runaway
stars than in stars belonging to  clusters or associations (Hubrig et
al.\ \cite{Hubrig2011b}; Hubrig et al.\ \cite{Hubrig2011a}). The
currently best available astrometric, spectroscopic, and photometric
data were used to calculate  the kinematical status of magnetic O-type
stars with previously unknown space velocities.  The results  suggest
that most of the magnetic O-type stars  can be considered as candidate
runaway stars.

The available observational material suggests that $\zeta$\,Oph is a
main sequence single star in the field with runaway characteristics.
Usually, to explain the origin of massive stars in the field, two
mechanisms are discussed in the literature.  In one scenario, close
multibody interactions in a dense cluster cause one or more stars to
be scattered out of the region (e.g.\ Leonard \& Duncan
\cite{LeonardDuncan1990}). For this mechanism, runaways are ejected in
dynamical three- or four-body interactions.  An alternative mechanism
involves a supernova (SN) explosion within a close binary, ejecting
the secondary due to the conservation of momentum (Zwicky
\cite{Zwicky1957}; Blaauw \cite{Blaauw1961}).  Blaauw
(\cite{Blaauw1952}) suggested the origin of $\zeta$\,Oph in the
Scorpius OB2 association due to its proper motion vector, which points
away from the association.  More recently, Hoogerwerf et
al.\ (\cite{Hoogerwerf2001}) suggested that the star gained it space
velocity of $\sim$30\,km\,s$^{-1}$ in a supernova explosion within a
close binary in Upper Scorpius about 1--2\,Myr ago.  The authors
identified PSR~B1929+10 as an associated pulsar with a characteristic
age of $\sim$3\,Myr, consistent with the kinematic age of $\zeta$\,Oph
within the uncertainties.  Tetzlaff et al.\ (\cite{Tetzlaff2010})
reinvestigated the scenario of a binary SN in Upper Scorpius involving
$\zeta$\,Oph and PSR~B1929+10 and concluded that it is very likely
that both objects were ejected during the same supernova event. In
their work, the considered association age range implies that the
progenitor star of the produced neutron star had a spectral type
between O6/O7 and O9 with a mass range from 18 to 37\,\Msun{}.  The
X-ray emission of the pulsar seems to be dominated by non-thermal
radiation processes (e.g.\ Becker et al.\ \cite{Becker2006}).  An
arc-like nebula surrounding PSR~B1929+10 and extending up to
10\arcsec{} was identified in {\em Chandra} data and interpreted as a
bow-shock nebula (Hui \& Becker \cite{HuiBecker2008}).
The estimated magnetic field strength in the shocked
region accounts for $\sim$75\,$\mu$G, while the typical magnetic field
strength in the ISM is about 2--6\,$\mu$G.

\begin{figure}
\centering
\includegraphics[width=0.45\textwidth]{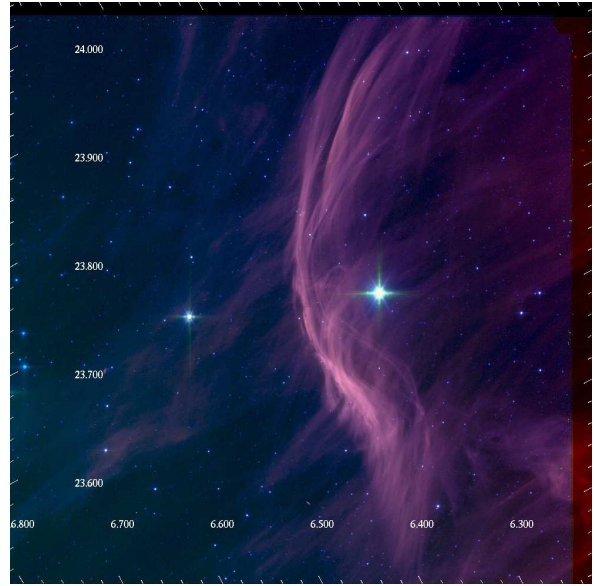}
\caption{(online colour at: www.an-journal.org) Combined IR {\em Spitzer} IRAC (3.6\,\mim\ blue,
  4.5\,\mim\ green, 8.0\,\mim\ red) image of the bow shock around the
  runaway star \zoph. Archival data have been used.
  Galactic coordinates are shown. The image size is $\sim 36'
  \times 31'$.  }
\label{fig:bsh}
\end{figure}

The presence of a bow-shock nebula has also been detected for $\zeta$\,Oph.
Figure~\ref{fig:bsh} shows an image based on archival {\em Spitzer}
IRAC maps (AOR 17774848). 
Recently, Kobulnicky \etal\ (\cite{Kobulnicky2010})
analyzed a sample of bow shocks around massive stars in Cygnus-X.
They used the analytical description of momentum-driven bow shocks  and
dust/polycyclic aromatic hydrocarbon emission models to estimate
stellar mass loss rates from the observed properties of the bow shocks. 
It was found that mass-loss rates in the range between
$10^{-7}$\myr\ and a few times $10^{-6}$\myr\ are required to
generate the bow shocks around typical B2V - O5V type stars. 

The mass-loss rate \mdot\ of this star was empirically obtained from
different diagnostics by a number of authors. Repolust
\etal\ (\cite{Repolust2004}) fitted the H$\alpha$ photospheric
  absorption line and derived the upper limit on the \zoph\ mass-loss rate
  as $1.8\times 10^{-7}$\myr.  Fullerton \etal\ (\cite{Fullerton2006})
  determined the radio-based mass-loss rate of \zoph\ as $1.1\times
  10^{-7}$\myr. The mass-loss rates determined from radio depend on
  the square of the density since the physical mechanism responsible for
  the radio emission is free-free emission.  On the other hand,
  Fullerton \etal\ derive a much smaller mass-loss rate from fitting the
  UV P\,{\sc v} resonance doublet, the product of the mass-loss rate
  and the ion fraction of P$^{+4}$ being only $\dot{M}q({\rm P^{+4}})
  \lsim 1.3 \times 10^{-10}$\myr\ with $q({\rm P^{+4}})\lsim 1$.  The
  mass-loss rates derived from fitting the wind profiles of UV
  resonance lines depend linearly on the density. To resolve this
  discordance in mass-loss determinations based on $\rho^2$- and
  $\rho$-diagnostics, Fullerton \etal\ suggest that the winds are strongly
  clumped with a volume filling factor of $\sim$10$^{-3}$--10$^{-5}$.
  Marcolino \etal\ (\cite{Marcolino2009}) analyzed optical and UV
  spectra of \zoph\ among their sample of O-type dwarfs. They derive
  an upper limit on the mass-loss rate of \zoph\ as $1.6 \times
  10^{-9}$\myr\ if the wind was smooth. This value agrees with the
  $\dot{M}q({\rm P^{+4}})$ value obtained by Fullerton
  \etal\ (\cite{Fullerton2006}).

Using the example of the O-type supergiant $\zeta$\,Puppis, Oskinova
\etal\ (\cite{Oskinova2007}) demonstrated that the discordance of
mass-loss rates found by Fullerton \etal\ can be overcome by
accounting for stellar wind porosity (see also Sundqvist
\etal\ \cite{Sundqvist2010}). It was found for the O5Ia star
$\zeta$\,Puppis that only a moderate reduction of the mass-loss rate
by a factor of 2--3 (compared to the smooth wind models) is required
to reproduce both H$\alpha$ and P\,{\sc v} lines. If this result
  holds also for non-supergiant O type stars, the mass-loss rate of
  \zoph\ is only a few times lower compared to the radio-based
  mass-loss determined by Fullerton \etal, i.e.\
$\sim$10$^{-7}$\,\myr. Importantly, this mass-loss rate is in agreement with values  
that are required to produce bow shocks around O stars 
(Kobulnicky \etal\ \cite{Kobulnicky2010}). 

An additional aspect, which may hint at the presence of a magnetic
field in runaway stars, is that a number of individual abundance
studies indicate nitrogen enrichment in the atmospheres of runaway
stars (e.g.\ Boyajian et al.\ \cite{Boyajian2005}).  Nitrogen
enrichment was found in $\zeta$\,Oph by Villamariz \& Herrero
(\cite{VillamarizHerrero2005}).  Recent NLTE abundance analyses (e.g.,
Morel et al.\ \cite{Morel2008}; Hunter et al.\ \cite{Hunter2008})
suggest that slow rotators have peculiar chemical enrichment such as
nitrogen excess or boron depletion, and these peculiarities are linked
to the presence of a magnetic field.
On the other hand, Hubrig et al.\ (\cite{Hubrig2011c}) showed that
some magnetic massive stars previously assumed to be slow rotators,
are in fact fast rotators, but are viewed close to their rotation
poles.

To test the magnetic nature of this particularly interesting rapidly
rotating runaway star, we acquired spectropolarimetric observations
with the low-resolution spectropolarimeter FORS\,1 at the VLT. In this
work we report the first detection of a magnetic field in this star.


\section{Magnetic field measurements}
\label{magnetic_field}

\begin{figure}
\centering
\includegraphics[width=0.45\textwidth]{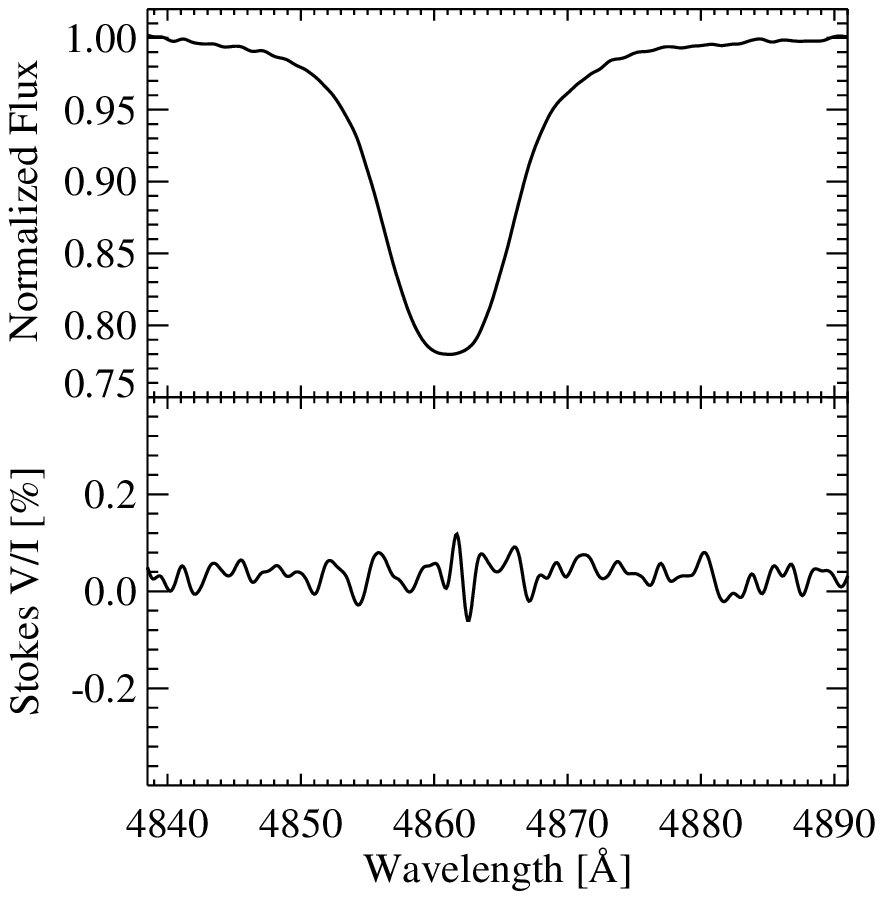}
\includegraphics[width=0.45\textwidth]{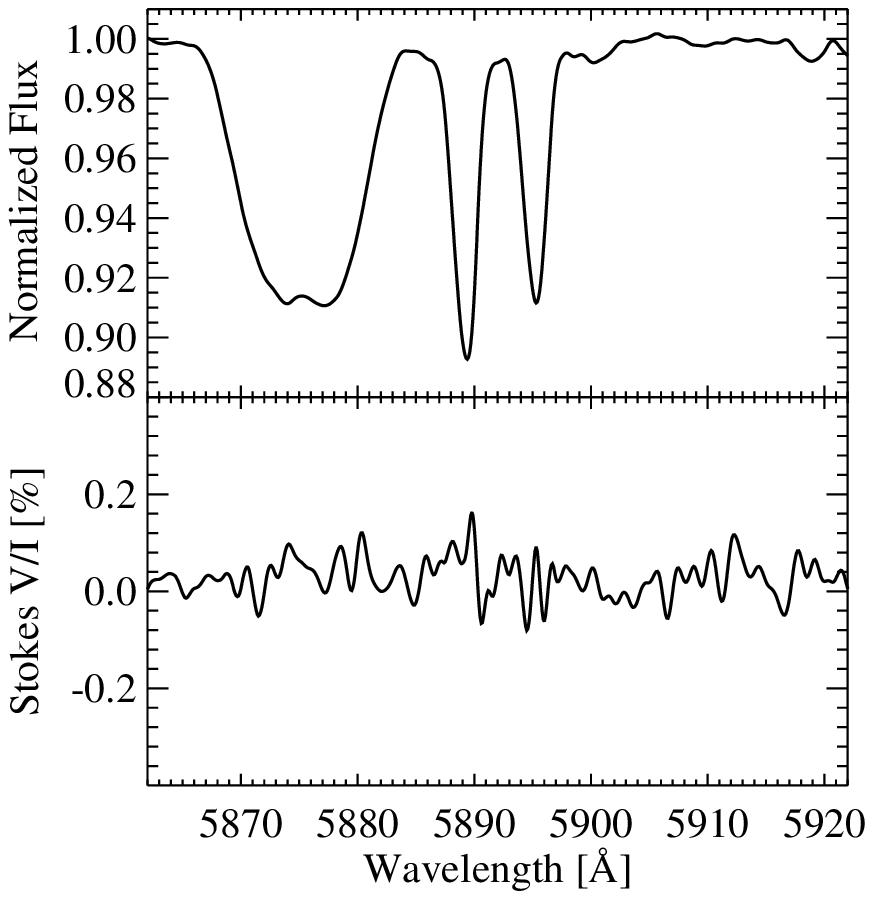}
\caption{
Stokes $I$ and $V$ spectra of $\zeta$\,Oph in the spectral regions around H$\beta$ (top) and the Na~I doublet
(bottom). 
}
\label{fig:a}
\end{figure}

Spectropolarimetric observations with FORS\,1 (Appenzeller et al.\ \cite{Appenzeller1998})
were obtained on 2008 May 23
(MJD\,54609.34) using grism 600B and a slit width of  0$\farcs$4 to achieve a spectral resolving power 
of $R\approx2000$.
The use of the mosaic detector made of 
blue optimized E2V chips and a  pixel size of 15\,$\mu$m allowed us to cover a large
spectral range, from 3250 to 6215\,\AA{}, which includes all hydrogen Balmer lines 
from H$\beta$ to the Balmer jump.
Six continuous series of two exposures with durations between 0.3 and 3\,sec were taken
at two retarder waveplate setups ($\alpha=+45^\circ$ and $-$45$^\circ$). For all but the first exposure pairs
we achieved a signal-to-noise ratio between 1000 and 1500.
More details on the observing technique with FORS\,1 can be 
found elsewhere (e.g.,  
Hubrig et al.\ \cite{Hubrig2004a}, \cite{Hubrig2004b}, and references therein).
The mean longitudinal 
magnetic field, $\left< B_{\rm z}\right>$, was derived using 

\begin{equation} 
\frac{V}{I} = -\frac{g_{\rm eff} e \lambda^2}{4\pi{}m_ec^2}\ \frac{1}{I}\ 
\frac{{\rm d}I}{{\rm d}\lambda} \left<B_{\rm z}\right>, 
\label{eqn:one} 
\end{equation} 

\noindent 
where $V$ is the Stokes parameter that measures the circular polarisation, $I$ 
is the intensity in the unpolarised spectrum, $g_{\rm eff}$ is the effective 
Land\'e factor, $e$ is the electron charge, $\lambda$ is the wavelength, $m_e$ the 
electron mass, $c$ the speed of light, ${{\rm d}I/{\rm d}\lambda}$ is the 
derivative of Stokes $I$, and $\left<B_{\rm z}\right>$ is the mean longitudinal magnetic 
field. 
The longitudinal magnetic field was measured in two ways: using only the absorption hydrogen Balmer 
lines or using the entire spectrum including all available absorption lines.
We obtain for the mean longitudinal magnetic field using all available absorption lines 
$\left< B_z\right>_{\rm all}= 141\pm45$\,G and for the mean longitudinal magnetic field using 
the hydrogen Balmer lines $\left< B_z\right>_{\rm hyd}= 123\pm54$\,G.
Our detection  using the entire spectrum has a significance of 
3.1$\sigma$, determined from the formal uncertainties we derive. 
In the Stokes $V$ spectra distinct Zeeman signatures are well visible at the position 
of hydrogen and metal lines. In Fig.~\ref{fig:a} we display Stokes $I$ and $V$ spectra
in the spectral regions around H$\beta$ and the Na~I doublet. 

\begin{figure}
\centering
\includegraphics[width=0.35\textwidth]{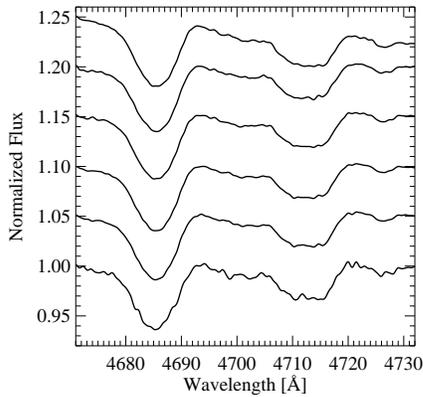}
\caption{
Six Stokes $I$ spectra corresponding to six sub-exposures in the region around He~II 4686\,\AA{}
and He~I 4713\,\AA{}. Small line profile variations in the He~I line are already detectable on the time
scale of only 13\,min.
}
\label{fig:x}
\end{figure}

In Fig.~\ref{fig:x} we present time series of Stokes $I$ spectra corresponding to 
our data set of six sub-exposures in the region around He~II 4686\,\AA{} and He~I 4713\,\AA{}.
Although the time lapse between the observations of the first pair and the last pair is only 13 minutes, 
some small line profile variations in the He I line are already detectable at such short time scales.  



The only other measurements of the magnetic field in $\zeta$\,Oph have been presented by Schnerr et al.\
(\cite{Schnerr2008}) who used the MuSiCoS spectropolarimeter to derive Stokes $I$ and $V$ spectra 
with the Least Square Deconvolution method. No longitudinal magnetic field was detected in this star
at a level more than 3$\sigma$. However, the measurement errors in their observations were in the range
from 700\,G to more than 3\,kG. 

Massive stars usually end their evolution with a final supernova explosion, producing neutron stars 
or black holes. The initial masses of these stars range from $\sim$8--10\,\Msun{} to 
100\,\Msun{} or more, which correspond to spectral types earlier than about B2. 
Contrary to the case of Sun-like stars, the
magnetic fields of stars on the upper main sequence (Ap/Bp stars),
white dwarfs, and neutron stars are dominated by large spatial
scales and do not change on yearly time scales. In each of these
classes there is a wide distribution of magnetic field strengths,
but the distribution of magnetic fluxes appears to be similar in
each class, with maxima of $\Phi_\mathrm{max}=\pi R^2B\sim 10^{27-28}\mathrm{G~cm^2}$
(Reisenegger \cite{Reisenegger2001};
Ferrario \& Wickramasinghe \cite{FerrarioWickramasinghe2005}), arguing for a fossil field
whose flux is conserved along the path of stellar evolution.
According to Reisenegger (\cite{Reisenegger2009}) the magnetic fluxes have possibly been generated on or 
even before the main-sequence
stage and then inherited by the compact remnants. 

The magnetic field strength of the pulsar PSR~B1929+10 is
0.5129$\times$10$^{12}$\,Gauss.
Assuming simple conservation of magnetic flux we obtain
a field strength of just a few Gauss for the more massive pulsar
progenitors.
For $\zeta$\,Oph, which is supposed to be formed in a binary SN, sharing
the same environment
with the SN progenitor, the expected field strength would be of the
order of 10\,G.
This value is notably lower than our current measurement, possibly
indicating that either the magnetic field
of this middle-aged pulsar has significantly decayed during the few Myrs
after the SN explosion or  other mechanisms
play a role in the generation of magnetic fields in O-type stars.


\section{Discussion}
\label{sect:discussion}


$\zeta$\,Oph has been extremely well studied in all wavelength ranges, from the X-ray by all major X-ray satellites
(with the exception of \xmm) to the infrared region with {\em Spitzer}.
In view of the detection of a magnetic field on \zoph\ reported
in this work, we review its X-ray properties with the aim to
understand whether the X-ray emission of \zoph\ is dominated by magnetic
or wind instability processes. 

Babel \& Montmerle (\cite{BabelMontmerle1997}) studied the case of a rotating star with a
dipole magnetic field sufficiently strong to confine stellar wind.
The magnetic field locally dominates the bulk motion of stellar wind,
when the ratio of magnetic to kinetic energy density, $B^2/\mu_0\rho
v^2 > 1$, where $v$ is the supersonic flow speed.  A collision between
the wind streams from the two hemispheres in the closed magnetosphere
leads to a strong shock and X-ray emission.

MHD simulations in the framework of this magnetically confined
wind shock (MCWS) model were performed by ud-Doula \& Owocki (\cite{udDoulaOwocki2002})
and Gagn{\'e} \etal\ (\cite{Gagne2005}). Using their notation, the wind is
confined when \mbox{$\eta_\ast\equiv
  (R_\ast^2B^2)(\dot{M}\vinf)^{-1} > 1$}.  New observations are
required to establish whether the magnetic field of \zoph\ is a
dipole. However, for the purpose of this discussion, let us assume that
the field has an average strength of 150\,G. Using the stellar
parameters of \zoph\ as inferred by Marcolino \etal\ (\cite{Marcolino2009}),
we estimate $\eta_\ast(\zoph)\sim 10^3$, i.e.\ the magnetic field should dominate
the wind motion up to the Alfv{\'e}n radius that is located at $\lsim$10\,$\Rstar$.
In this case, the X-ray emission should mainly originate from the MCWS.

The MCWS model predicts that the X-ray emitting plasma should be
located at a few \Rstar\ from the photosphere; that the X-ray emission
lines should be narrow; that the X-ray luminosity should be higher and
the spectrum harder than in non-magnetic stars; that in 
  oblique magnetic rotators the X-ray emission should be modulated
periodically as a consequence of the occultation of the hot plasma by
a cool torus of matter, or by the opaque stellar core.

\begin{figure}
\centering
\includegraphics[width=\columnwidth]{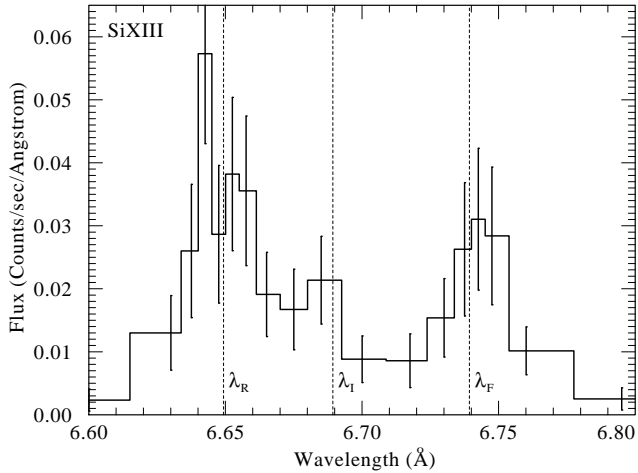}
\caption{The Si\,{\sc xiii} line observed in the spectrum of \zoph{}
(co-added MEG $\pm$1). Vertical dashed lines indicate the rest-frame
wavelength: $\lambda_{\rm R}$ -- resonant line, $\lambda_{\rm I}$ -- 
sum of intercombination lines, $\lambda_{\rm F}$ -- forbidden line. 
The rest-frame wavelengths are corrected for the radial velocity 
taken from Hoogerwerf \etal\ (\cite{Hoogerwerf2001}).}
\label{fig:sixiii}
\end{figure}

The lines of He-like ions observed in X-ray spectra are useful to
derive the location of the line formation region in hot stars 
  because forbidden line emission is depressed by ultraviolet
  pumping. The latter depends on the distance to the stellar
  photosphere (Gabriel \& Jordan \cite{GabrielJordan1969}; Blumenthal
\etal{} \cite{Blumenthal1972}).  The Si\,{\sc xiii} line observed in the
\cxo\ HETGS/MEG spectrum is shown in Fig.~\ref{fig:sixiii}.  
 prominent forbidden line can easily be distinguished in this fugure, while
normally forbidden lines are strongly suppressed in the spectra of
OB-type stars.  The presence of the forbidden line implies that the
line formation region is located far from the photosphere, so that the
radiative excitation does not lead to the depopulation of the
corresponding metastable energy levels. Waldron \& Cassinelli
(\cite{WaldronCassinelli2008}) found that the Si\,{\sc xiii} line is
formed at $1.8 \pm 0.7$\,\Rstar\ in \zoph\ and that other He-like
lines are formed even further out in the wind. Interestingly, the
  strong forbidden Si\,{\sc xiii} line is also observed in the {\em
    Chandra} spectrum of the magnetic star $\tau$\,Sco. Cohen
  \etal\ (\cite{Cohen2003}) derive a Si\,{\sc xiii} line formaiton
  radius for $\tau$\,Sco in the range between 1.1\,$R_\ast$ and 1.5\,$R_\ast$.
These radii of line formation are smaller than those found in the
  prototypical MCWS model object $\theta^1$\,Ori\,C, $3.4 \pm
  0.8$\,\Rstar\ (Waldron \& Cassinelli \cite{WaldronCassinelli2008}).

\begin{figure}
\centering
\includegraphics[width=\columnwidth]{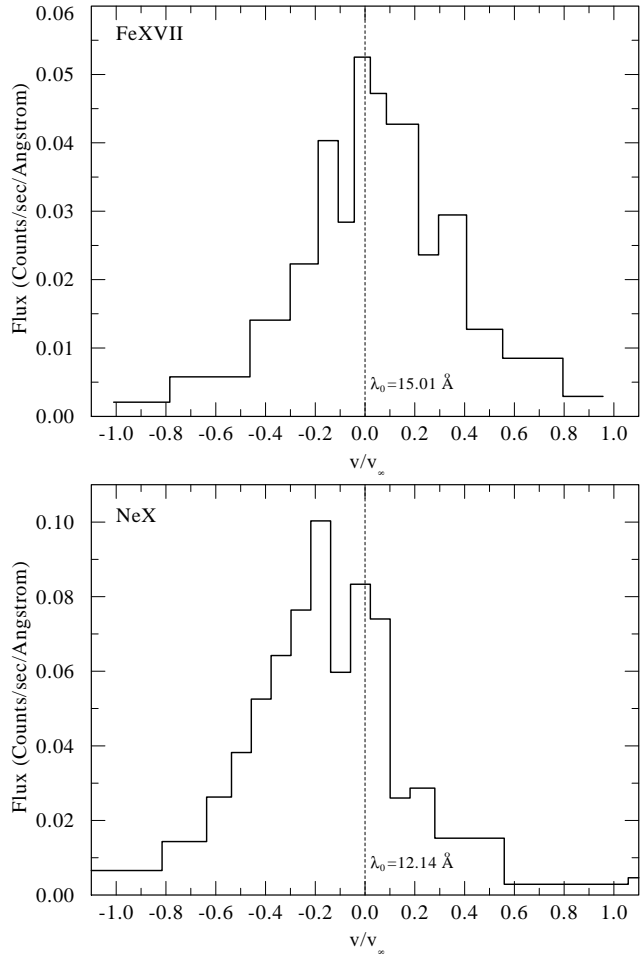}
\caption{Fe\,{\sc xvii} (upper panel) and Ne\,{\sc x} (lower panel)
  lines observed in the spectrum of \zoph\ (co-added MEG $\pm$1). 
Vertical dashed lines indicate the rest-frame wavelength, corrected for the 
radial velocity. }
\label{fig:vel}
\end{figure}

Oskinova \etal\ (\cite{Oskinova2006}) studied the \cxo\ spectrum of \zoph\ among
other O-type stars. They found that the X-ray emission lines in this
star are narrow and that the signatures of wind absorption on line
profiles are weak.  Figure~\ref{fig:vel} shows the Fe\,{\sc xvii} and 
Ne\,{\sc x} lines as measured by \cxo\ plotted over units of the wind
terminal  velocity, \vinf=1550\,km\,s$^{-1}$. The lines are only slightly broadened, 
if at all. 

The X-ray luminosity of \zoph, $\Lx=1.2\times 10^{31}$\,erg\,s$^{-1}$
and the ratio $\Lx/\Lbol=4\times 10^{-8}$ are quite usual among late
type OV stars (Oskinova \etal\ \cite{Oskinova2006}). Adopting the
  mass-loss rate from \zoph\ as $\lsim 1.8\times 10^{-7}$\,\myr{},
  Oskinova \etal{} (\cite{Oskinova2006}) noticed that in \zoph{} the ratio
  of X-ray to wind mechanical luminosity $L_{\rm mech}$ ($\mdot\vinf^2/2$), $L_{\rm
    X}/L_{\rm mech} \msim 8.5\times 10^{-5}$, is a few times
    higher than in other single O-type stars. This may be related to
  the lower wind opacity in \zoph, or it may hint at some additional
  mechanism of X-ray generation besides the intrinsic wind shocks.

From their analysis of \cxo\ spectra, Zhekov\& Palla
(\cite{ZhekovPalla2007}) derived the differential emission measure (DEM)
for \zoph\ among other OB stars in their sample. They found that in
\zoph\ the DEM sharply peaks at about 6\,MK. While this is a
significantly lower temperature than found for the DEM peak in case of
$\theta^1$\,Ori\,C (50\,MK), it is higher than found for other stars of
similar spectral types ($\sim$3\,MK). Thus, considering the X-ray
temperature of \zoph{}, it is not straightforward to attribute its X-ray
emission to the MCWS. On the other hand, recent studies of O stars with
detected magnetic fields (e.g., HD\,191612, HD\,108) show that their
X-ray properties are diverse (Naz{\'e} \etal\ \cite{Naze2004},
\cite{Naze2010b}) and may be difficult to fully reconcile with the
predictions of the MCWS model. 

Clearly, new observations are needed to  better understand the magnetic field of \zoph\
and its  link with the X-ray emission from this star.

{
\acknowledgements
We would like to thank Y.~Naz\'e for drawing our attention to this interesting star and the 
anonymous referee for valuable comments.
LMO acknowledges the financial support from grant number  FKZ~50~OR~1101. This work used 
archival data obtained with the Spitzer Space Telescope, which is operated by the Jet 
Propulsion Laboratory, California Institute of Technology, under a contract with NASA. 
We also used data obtained from the Chandra Data Archive and software provided
by the Chandra X-ray Center (CXC).}


\label{lastpage}

\end{document}